\documentclass[preprint,12pt]{elsarticle}
\pdfoutput=1
\usepackage{hyperref}

\usepackage{framed}

\usepackage{amssymb}
\usepackage{amsmath}
\usepackage{enumitem}
\usepackage{graphicx
}
\usepackage{url}
\usepackage{listings}
\usepackage{bm}
\usepackage{multirow}
\usepackage{array}
\usepackage{color}
\usepackage{threeparttable}
\usepackage{float}
\usepackage{wrapfig}
\usepackage{xcolor}
\usepackage{tikz}
\usepackage{makecell}
\usepackage{cuted}
\usepackage{capt-of}
\definecolor{shadecolor}{rgb}{0.95,0.95,0.92}
\usepackage{xcolor}
\usepackage{xcolor,colortbl}
\usepackage{listings}

\definecolor{shadecolor}{rgb}{0.95,0.95,0.92}
\definecolor{codegreen}{rgb}{0,0.6,0}
\definecolor{codegray}{rgb}{0.5,0.5,0.5}
\definecolor{codepurple}{rgb}{0.58,0,0.82}
\definecolor{backcolour}{rgb}{0.95,0.95,0.92}

\lstdefinestyle{mystyle}{
    commentstyle=\color{codegreen},
    keywordstyle=\color{magenta},
    numberstyle=\tiny\color{codegray},
    stringstyle=\color{codepurple},
    basicstyle=\ttfamily\footnotesize,
    breakatwhitespace=false,
    breaklines=true,
    captionpos=b,
    keepspaces=true,
    numbers=left,
    numbersep=5pt,
    showspaces=false,
    showstringspaces=false,
    showtabs=false,
    tabsize=2
}

\newcommand\crule[3][black]{\textcolor{#1}{\rule{#2}{#3}}}
\usetikzlibrary{shapes,arrows,arrows,positioning,matrix}
\urldef{\mailsc}\path|matliuj@nus.edu.sg|

\journal{Journal of \LaTeX\ Templates}

\begin{document}

\begin{frontmatter}

\title{Rethinking Medical Image Reconstruction via Shape Prior, Going Deeper and Faster: Deep Joint Indirect Registration and Reconstruction }

%


\author{Jiulong Liu$^{1}$, Angelica I Aviles-Rivero$^{2}$,  Hui Ji$^{1}$ and  Carola-Bibiane Sch\"{o}nlieb$^{3}$}
\address{$^{1}$ Department of Mathematics, National University of Singapore (NUS) \\
$^{2}$Department of Pure Mathematics and Mathematical Statistics and $^{3}$Department of Applied Mathematics and Theoretical Physics, University of Cambridge, UK
}

\begin{abstract}
Indirect image registration is a promising technique to improve image reconstruction quality by providing a shape prior for the reconstruction task. In this paper, we propose a novel hybrid method that seeks to reconstruct high quality images from few measurements whilst requiring low computational cost. With this purpose, our framework  intertwines indirect registration and reconstruction tasks is a single functional.  It is based on two major novelties. Firstly, we introduce a model based on deep nets to solve the indirect registration problem, in which the inversion and registration mappings are recurrently connected through a fixed-point interaction based sparse optimisation. Secondly, we introduce specific inversion blocks, that use the explicit physical forward operator, to map the acquired measurements to the image reconstruction. We also introduce registration blocks based deep nets to predict the registration parameters and warp transformation accurately and efficiently.  We demonstrate, through extensive numerical and visual experiments, that our framework outperforms significantly classic reconstruction schemes and other bi-task method; this in terms of both  image quality and computational time. Finally, we show generalisation capabilities of our approach by demonstrating their performance on  fast Magnetic Resonance Imaging (MRI), sparse view computed tomography (CT) and low dose CT with measurements much below the Nyquist limit.

\end{abstract}


\end{frontmatter}

\section{Introduction}
Image reconstruction and registration are two fundamental tasks in medical imaging. They are necessary to gain better insights in different  applications - including  diagnostic,  surgery planning and radiotherapy (e.g.~\cite{alp1998head,wein2008automatic,crum2004non,smit2016pelvis}) just to mention few. For several medical imaging modalities, for example Magnetic Resonance Imaging (MRI),  it is highly desirable to reduce the number of the acquired measurements to avoid image degradation~\cite{sachs1995diminishing,zaitsev2015motion} (for example - geometric distortions and blurring effects). This with the purpose to deal with the central problem in MRI - the long acquisition time. However, to perform these tasks from  undersampled and  highly corrupted measurements become even a more challenging problem yet of great interest from the theoretical and practical points of view.

There have been different attempts to perform image reconstruction and registration in the community, which these two tasks are performed either separately and most recently jointly.  For image reconstruction the majority of algorithmic approaches follow the notion of Compressed Sensing (CS)- e.g. \cite{lustig2007sparse,liang2007spatiotemporal,lingala2011accelerated,Otazo::2015,zhang2015accelerating}. Most recently, there has been a growing interest in exploring similarity of image structures of to-be-registrated images as shape prior e.g.~\cite{liu20155d}, and  deep learning based reconstruction approaches e.g.~\cite{sun2016deep,hyun2018deep,hammernik2018learning}. For a detailed survey in image reconstruction, we refer the reader to~\cite{ravishankar2019image}.

Whilst for  image registration, that seeks to find a mapping  that aligns two or more images, the body of literature has reported promising results. These can be roughly divided in rigid and deformable algorithmic approaches. Whilst rigid registration, e.g.~\cite{adluru2006model,wong2008first,johansson2018rigid}, has shown promising results, it is not enough robust to describe complex physiological motions. Deformable registration offers greater opportunities to describe complex motion - for example~\cite{beg2005computing,cao2005large,vercauteren2009diffeomorphic}. We refer the reader to~\cite{sotiras2013deformable} for an extensive revision on deformable registration. More recently, deformable image registration has also benefited of the potentials of deep learning- e.g.~\cite{yang2017quicksilver,shen2019networks,balakrishnan2019voxelmorph,haskins2019deep}. However,  these approaches assume that the given images are already reconstructed.

A commonality of the aforementioned approaches is that they perform the reconstruction and registration tasks separately. In very recent developments in the area, e.g.~\cite{aviles2018compressed,corona2019variational}, have shown that performing those tasks jointly can reduce error propagation resulting in improving accuracy whilst achieving better generalisation capabilities~\cite{caruana1997multitask}. However, a major bottleneck of such joint models is the computational complexity as they often seek to solve highly non-convex optimisation problems. Motivated by  the current  drawbacks in the literature, we address the problem of $-$ how to get higher quality reconstructed and registered images from noisy and undersampled MRI measurements whilst demanding low computational cost.


In this work, we address the previous question by proposing a new framework for simultaneous reconstruction and registration from  corrupted and undersampled MRI data. Our approach is framed as a deep joint model, in which \textit{these two task are intertwined in a single  optimisation model}. It benefits from the  theoretical guarantees of large deformation diffeomorphic metric mapping (LDDMM) and the powerful performance of deep learning.
Our modelling hypothesis is that by providing a shape prior (i.e. registration task) to the reconstruction task, one can boost the overall performance of the final reconstruction. Most precisely, our framework seeks to learn a network parametrised mapping $(u,g)\rightarrow f$, where $u$ is the image to be reconstructed and $g,f$ are the template and target images to-be-register.

We remark to the reader that unlike the works of that~\cite{yang2017quicksilver,shen2019networks,balakrishnan2019voxelmorph,haskins2019deep}, our approach follows a different philosophy which is based on three major differences. Firstly, we address the problem of \textit{indirect registration, in which the target image  is unknown but encoded in the indirect corrupted measurements (i.e. raw data)}. Secondly, \textit{our ultimate goal is to improve the final image reconstruction through shape prior (i.e. registration task)} instead of evaluate the tasks separately. Thirdly, unlike the work of that~\cite{lang2018template} we gain further computational efficiency and reconstruction quality through our registration blocks based deep nets.

We highlight that computing image reconstruction and indirect registration \textit{simultaneously} is even more challenging than performing the reconstruction and registration separately. This is because  $u$ is not explicitly given and is encoded in a corrupted measurement, and the general physical forward operators (e.g. Fourier and Radon transforms) are not trivial to be learnt~\cite{zhu2018image}. Therefore, \textit{to build  an end-to-end parameterised mapping for inverse problems is not straightforward via standard deep nets}. Motivated by the existing shortcomings in the body of literature, in this work we propose a novel framework, that to the best of our knowledge, it is the first hybrid method (i.e. a combination of a model-based and deep-learning based approaches) that intertwines reconstruction and indirect registration. Although we emphasise the application of fast MRI, we also show generalisation capabilities using Computerised  Tomography (CT) data. Whilst this is an relevant part of our approach, our contributions are:

\noindent
\begin{itemize}[noitemsep]
    \item We propose a novel mathematically well-motivated and computationally tractable framework for simultaneous reconstruction and indirect registration, in which we highlight: 
        \begin{itemize}
            \item  A framework based on deep nets for solving indirect registration efficiently, in which the inversion and registration mappings are recurrently connected through a fixed-point iteration based sparse optimisation.
            \item We introduce two types of blocks for efficient numerical solution of our bi-task framework. The first ones are specific inversion blocks that use the explicit physical forward operator, to map the acquired measurements to the image reconstruction. Whilst the second ones are registration blocks based deep nets to predict the registration parameters
            and warping transformation.
        \end{itemize}
    \item We exhaustively evaluate our framework with a range of numerical results and for several applications including fast MRI, sparse view computerised  tomography (CT) and low dose CT.
    \item We show that the carefully selected components in our framework mitigate major drawbacks of the traditional reconstruction algorithms resulting in significant increase in image quality  whilst decreasing substantially the  computational cost.
\end{itemize}



\section{When Reconstruction Meets LDDMM: A Joint Model}
In this section, we first introduce the tasks of image reconstruction and registration separately, and then, we describe \textit{how these two tasks can be cast in a unified framework}.

\medskip
Mathematically, the task of reconstructing a medical image modality, $u$, from a set of measurements $y$ reads:
\begin{equation}
  y= A u+\eta,
  \label{eq:recons}
\end{equation}
where $A$ is the forward operator associated with the acquired measurement $y$; and $\eta$ is the inherent noise. To deal with the ill-posedness of~\eqref{eq:recons}, one can be casted it as a variational approach as: $\text{argmin}_u \mathcal{D}(Au,y)+\alpha\mathcal{J}(u)$, where $\mathcal{D}$ is the data fidelity term, d $\mathcal{J}$ is a regularisation term to restrict the space of solutions, and $\alpha$ is a positve parameter balancing the influence of both terms.
Whilst the task of registering a template image, $g$, to a target one, $f$, can be cast as an optimisation  problem, which functional can be expressed as:
\begin{equation}
  E(\Phi)=R(\phi)+\frac{1}{\sigma}\|f \circ \phi^{-1}-g\|_2^2,
  \label{eq:reg}
\end{equation}

\noindent
where $\phi$ denotes a deformation map and $R(\phi)$ regularises the deformation map. In general, the registration problem is  ill-posed, and a regulariser, $R(\phi)$, is necessary to obtain a reliable solution. 
There are several methods proposed in the literature to regularise the deformation mapping~\cite{sotiras2013deformable}. One well-established algorithmic approach, due to its desirable mathematical properties, is  Large Deformation Diffeomorphic Metric Mapping (LDDMM)~\cite{cao2005large}.

In the LDDMM setting, the deformation map $\phi$ is assumed to be invertible (to make the deformation physically meaningful), and both $\phi$ and $\phi^{-1}$ should be sufficiently smooth, i.e. $\phi\in \operatorname{Diff}^{p}\left(\mathbb{R}^{n}\right)$, which is defined as:
\begin{equation}
\operatorname{Diff}^{p}\left(\mathbb{R}^{n}\right) :=\left\{\phi \in {C}^{p}\left(\mathbb{R}^{n}, \mathbb{R}^{n}\right) : \phi \text { is bijective with } \phi^{-1} \in {C}^{p}\left(\mathbb{R}^{n}, \mathbb{R}^{n}\right)\right\}
\end{equation}

The $\operatorname{Diff}^{p}\left(\mathbb{R}^{n}\right)$ forms a group with the identity mapping $\mathcal{I}$ as the neutral element. When small perturbations $\epsilon v$ of the identity mapping are applied to $\phi_{i-1}$,  at a particular time point $i-1$,  the deformation at the next time point $i$ becomes $\phi_{i}= (\mathcal{I}+\epsilon v)\circ \phi_{i-1}$, which can be described by the following difference equation:
\begin{equation}
  \frac{\phi_{i}-\phi_{i-1}}{\epsilon}=  v \circ \phi_{i-1},
\end{equation}
and leads to a continuous-time flow equation, which reads:
\begin{equation}
\phi_t(x,t)=  v(\phi(x,t),t).
\end{equation}

LDDMM is a PDE constrained optimisation problem, which can be formulated as:
\begin{equation}
\label{lddmm}
\left \{
\begin{array}{l}
  \min\limits_{ v} \gamma\int_{0}^{1}\|v\|_L^2+\frac{1}{2}\|f\circ \phi^{-1}(x,1)-g(x)\|_2^2\\
  s.t. ~ \phi_t(x,t)=v(\phi(x,t),t), \phi(x,0)=\mathcal{I}, \mbox{for}~ t\in [0,1]
\end{array}
\right .
\end{equation}
where $\|v\|_L^2=<Lv,v>, L$ is a self-adjoint differential operator, whose numerical solution  can be given via Euler-Lagrange equations~\cite{beg2005computing}.
Let the momentum $m$ be the dual of velocity, i.e. $m:=Lv$, and  $K$  the inverse of $L$ then \eqref{lddmm} can be expressed as a function of the momentum $m$ as:

\begin{equation}
\label{lddmm_momentum}
\left \{
\begin{array}{l}
  \min\limits_{ m(x,t)} \frac{\gamma}{2}\langle m(x,t), K m(x,t) \rangle+\frac{1}{2}\|f\circ \phi^{-1}(x,1)-g(x)\|_2^2\\
  s.t. ~~
\left\{
\begin{array}{l}
   \phi_t(x,t)=v(\phi(x,t),t)\\
   \phi(x,0)=\mathcal{I} \\
   m(x,t)-L v(x,t)=0 .
\end{array}
\right . \\
\end{array}
\right .
\end{equation}

From an optimisation point of view, instead of solving \eqref{lddmm} over all possible velocities $v$, one can apply the shooting formulation~\cite{vialard2012diffeomorphic} and account only for those  with least norm for a given $\phi$. Now when computing Euler-Lagrange equation to the regularisation term $\langle m(x,t), K m(x,t) \rangle$, one can get the Euler-Poincar\'{e} equation~\cite{holm1998eulerpoincar}:
\begin{equation}
     m_t(x,t)+ad^*_{v} m(x,t)=0,
\end{equation}

\noindent
where the adjoint action $ad_{v} u = d v\cdot u- d u\cdot v$ and the conjoint actions $ad^*_{v} $ is defined via $\langle ad^*_{v} m, u\rangle=\langle m, ad_{v} u\rangle$. Therefore, \eqref{lddmm_momentum} can be efficiently optimised  over $m(x,t)$  via Geodesic shooting. It can now be expressed as:
\begin{equation}
\label{lddmm_shooting}
\left \{
\begin{array}{l}
  \min\limits_{ m(x,0)} \frac{\gamma}{2}\langle m(x,0), K m(x,0) \rangle+\frac{1}{2}\|f\circ \phi^{-1}(x,1)-g(x)\|_2^2\\
  s.t. ~~
\left\{
\begin{array}{l}
   \phi_t(x,t)=v(\phi(x,t),t)\\
   \phi(x,0)=\mathcal{I} \\
   m(x,t)-L v(x,t)=0 \\
   m_t(x,t)+ad^*_{v} m(x,t)=0\\
\end{array}
\right . \\
\end{array}
\right .
\end{equation}

As we are interested in performing simultaneously reconstruction and registration. We now turn to describe how these two task can be intertwined in an unified framework.  Consider the target image $u$ to be encoded in a set of  measurements $y$, then one can join these two tasks, i.e. \eqref{eq:recons} and \eqref{eq:reg}, as a \textit{single optimisation problem}, which reads:

\begin{equation}
E(\Phi)=R(\Phi)+\frac{1}{\lambda}\|A u-y\|_2^2+\frac{1}{\sigma}\|u\circ \phi^{-1}-g\|_2^2
    \label{eq:jointM}
\end{equation}

One can naturally rewrite~\eqref{eq:jointM} using LDDMM via geodesic shooting ~\eqref{lddmm_shooting}. 
This results in the following expression:

\begin{equation}
\label{lddmm_indir_shooting}
\left \{
\begin{array}{l}
  \min\limits_{ m(x,0),u} \frac{\gamma}{2}\langle m(x,0), K m(x,0) \rangle+\frac{1}{2}\|A u(x)-y\|_2^2+\frac{\mu}{2}\|u\circ \phi^{-1}(x,1)-g(x)\|_2^2\\
  s.t. ~~
\left\{
    \begin{array}{l}
       \phi_t(x,t)=v(\phi(x,t),t)\\
       \phi(x,0)=\mathcal{I} \\
       m(x,t)-L v(x,t)=0 \\
       m_t(x,t)+ad^*_{v} m(x,t)=0.\\
    \end{array}
\right .
\end{array}
\right .
\end{equation}
where $K$ is the inverse of $L$. However, a potential shortcoming of \eqref{lddmm_indir_shooting}  is that the solution, via Euler-Lagrange method, is computationally expensive. In the next section, we describe how ~\eqref{lddmm_indir_shooting} can be efficiently solved  by using  Deep Learning. In particular, using deep nets parametrised  Douglas-Rachford iteration~\cite{lions1979splitting}.

\begin{figure}[t!]
\centering
\includegraphics[width=1\textwidth]{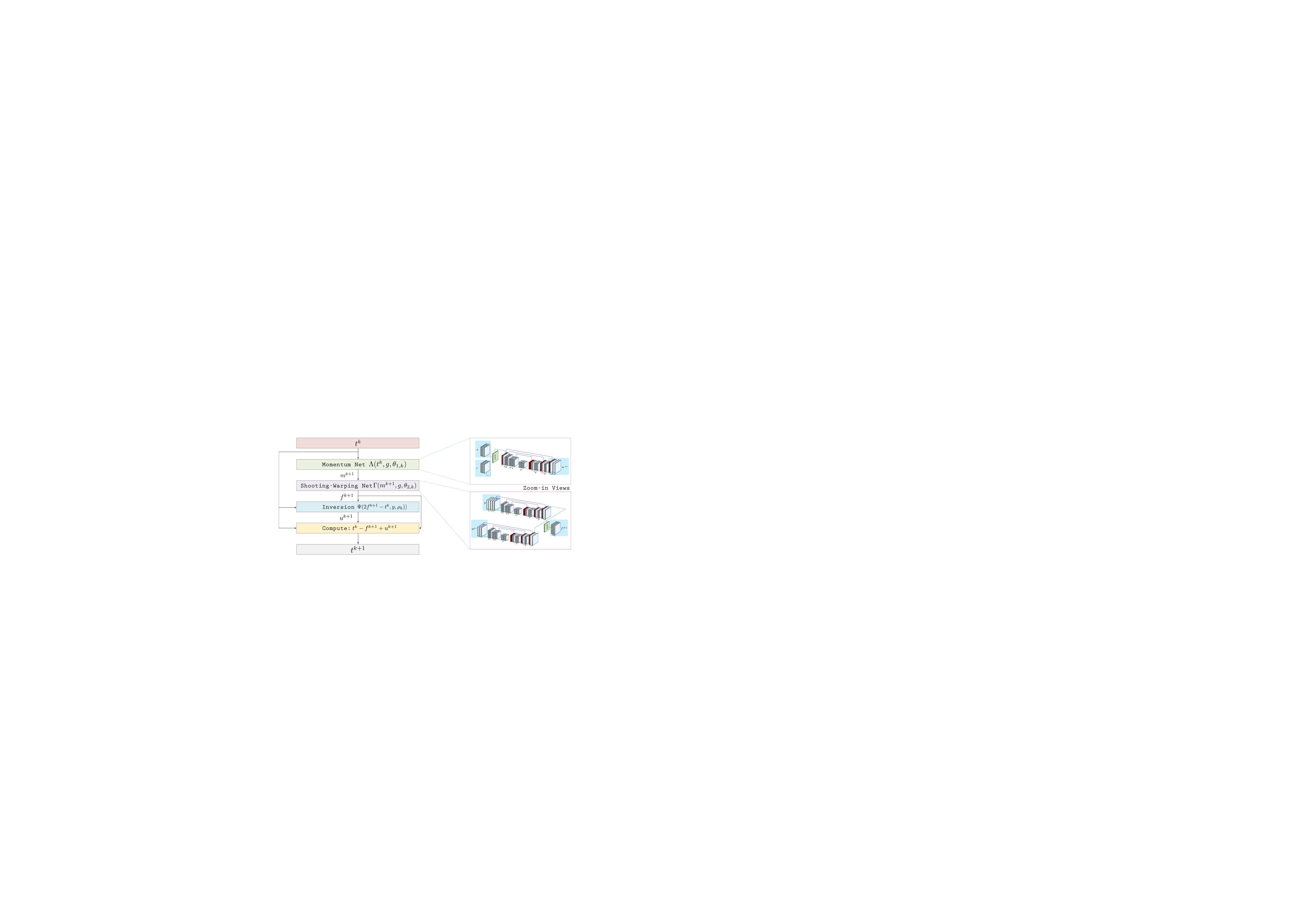}
\caption{\label{fig:teaser} Workflow of our proposed framework, in which the simultaneous reconstruction and registration is achieved using deep nets parametrised Douglas-Rachford iteration with $k$ stages ($k=0, 1, \cdots, N-1$) where the $t^0$ is initialized by $ u^0$ which can be reconstructed by a conventional method such as total variation regularised reconstruction. }
\end{figure}

\section{Deep Nets Paramatrised Douglas-Rachford Fixed-point Iteration of Sparsity Optimization (SOFPI-DR-Net) for Simultaneous Reconstruction and Registration}
In this section, we describe in details our novel framework that joins two tasks in a unified optimisation problem. We then demonstrate that it can be solved efficiently by splitting our optimisation model into more tractable sub-problems. We also define our inversion and registration blocks based on deep nets. Fig.~\ref{fig:teaser} displays the overview of our proposed frameworks.

\medskip
We remind to the reader that we seek to solve~\eqref{lddmm_indir_shooting} in a   computational tractable manner. The model \eqref{lddmm_indir_shooting} is equivalent to:

\begin{equation}
\label{lddmm_inshooting2}
\left \{
\begin{array}{l}
  \min\limits_{ m(x,0),u} \frac{\gamma}{2}\langle m(x,0), K m(x,0) \rangle+\frac{1}{2}\|A u-y\|_2^2+\frac{\mu}{2}\|f\circ \phi^{-1}(x,1)-g(x)\|_2^2\\
  s.t. ~~
\left\{
    \begin{array}{l}
      \phi_t(x,t)=v(\phi(x,t),t)\\
       \phi(x,0)=\mathcal{I} \\
       m(x,t)-L v(x,t)=0 \\
       m_t(x,t)+ad^*_{v} m(x,t)=0\\
       f=u.\\
      \end{array}
  \right .
\end{array}
\right.
\end{equation}

\noindent
An efficient manner to solve ~\eqref{lddmm_inshooting2} is via Alternating Direction Method of Multipliers (ADMM)/ Douglas-Rachford splitting, in which  one can break ~\eqref{lddmm_inshooting2} into more computational tractable sub-problems. Therefore, we solve ~\eqref{lddmm_inshooting2} via alternating minimisation, which yields to the following sub-problems:

\begin{equation}
\label{lddmm_admm}
\left \{
\begin{array}{l}
[\tikz\draw[red,fill=red] (0,0) circle (.5ex);] \hspace{0.2cm} u^{k+1}=\arg\min\limits_{u}\|A u-y\|_2^2+\frac{\rho}{2}\| u-f^k+b^k\|_2^2 \\
\left \{
\begin{array}{l}
\begin{split}
 [\tikz\draw[blue,fill=blue] (0,0) circle (.5ex);] \hspace{0.2cm} (f^{k+1}, m^{k+1})=&\arg\min\limits_{f, m(x,0)} \frac{\gamma}{2}\langle m(x,0), K m(x,0) \rangle\\
    &+\frac{\mu}{2}\|f\circ \phi^{-1}(x,1)-g\|_2^2+\frac{\rho}{2}\| u^{k+1}-f+b^k\|_2^2\\
 \end{split}\\
  s.t.
  \left\{
  \begin{array}{l}
       \phi_t(x,t)=v(\phi(x,t),t)\\
   \phi(x,0)=\mathcal{I} \\
   m(x,t)-L v(x,t)=0 \\
   m_t(x,t)+ad^*_{v} m(x,t)=0\\
    \end{array}
    \right .\\
\end{array}
\right .\\ \\
 b^{k+1}=b^k+(u^{k+1}-f^{k+1})
\end{array}
\right .
\end{equation}


\noindent
We now turn to give more details on the solution of each sub-problem. The first sub-problem [\tikz\draw[red,fill=red] (0,0) circle (.5ex);] can be solved by a general inversion method such as conjugate method as:
\begin{equation}
\label{denote_Psi}
u^{k+1}:=\Psi(f^k-b^k,y)
\end{equation}

However, solving the second sub-problem  [\tikz\draw[blue,fill=blue] (0,0) circle (.5ex);]   is similar to LDDMM, and therefore, solving it is still computationally expensive. The solution is denoted as:
\begin{equation}
\label{denote_Phi}
f^{k+1}:=\Phi(u^{k+1}+b^k,g).
\end{equation}

The problem  \eqref{lddmm_admm}  can be also rewritten as a fixed-point iteration as:
\begin{equation}
\label{updated1}
t^{k+1}=b^k+ u^{k+1}
\end{equation}

\noindent
and then one can obtain:
\begin{equation}
\label{updated2}
f^{k}=\Phi(t^k,g),
\end{equation}
and
\begin{equation}
\label{updateb2}
\begin{array}{rl}
  {b}^{k}&={b}^{k-1}+({u}^{k}-{f}^{k})\\
            &={b}^{k-1}+t^k-b^{k-1}- f^{k}\\
            &=t^k- f^{k}\\
            &=t^k-\Phi(t^k,g).\\
\end{array}
\end{equation}

Based on the update of $u^{k+1}$ along with  \eqref{updated1},  \eqref{updated2} and \eqref{updateb2}, a  fixed-point iteration for \eqref{lddmm_admm} reads:

\begin{equation}
\label{eqn:fixpit}
\begin{array}{rl}
  t^{k+1}&=b^k+ u^{k+1}\\
  &=b^k+ \Psi({f}^{k}-{b}^k, y)\\
  &=t^k-\Phi(t^k,g, \theta_k)+\Psi(2\Phi(t^k,g)-{t}^k, y)) .\\
\end{array}
\end{equation}

The fixed-point iteration is also called Douglas-Rachford iteration \cite{lions1979splitting}. We consider parameterise the inversion mapping $\Psi$ and  registration mapping $\Phi$  for the Douglas-Rachford iteration (\ref{eqn:fixpit}). For $\Psi$, a learnable inversion $\Psi(v, y, \rho)$ - with the parameter $\rho$ in optimisation  model \eqref{lddmm_admm} considered to be either learnable or manually tunable - is used in the  fixed-point iteration of (\ref{eqn:fixpit}). Whilst for the  registration mapping,  $\Phi$,  a parameterised  $\Phi(t,g,\theta)$ is  replaced in the the fixed-point iteration \eqref{eqn:fixpit}.  To use LDDMM framework to regularise the registration parameters,  we use  $\Phi(u,g,\theta)$ consisting of a momentum prediction neural net $m=\Lambda(t,g,\theta_{1})$ instead of searching momentum by \eqref{lddmm_indir_shooting}. Moreover,
a shooting-warping neural net $f=\Gamma(m,g,\theta_{2})$, which mimics the shooting and warping in \eqref{lddmm_indir_shooting}, is used. Finally, our framework for  parameterising the algorithm (\ref{lddmm_indir_shooting}) with $N$ stages is obtained by computing:

\begin{equation}
\label{eqn:fixpit-net}
\begin{aligned}
t^{k+1}&=t^k-\Gamma(\Lambda(t^k,g,\theta_{1,k}),g,\theta_{2,k})+\Psi(2\Gamma(\Lambda(t^k,g,\theta_{1,k}),g,\theta_{2,k})-{t}^k, y, \rho_k)),
\end{aligned}
\end{equation}

\noindent
for $k=0, 1, \cdots, N-1$.  We now give more details on the Deep Nets used for $\Psi$, $\Lambda$ and $\Gamma$ in each stage.

\subsection{The Inversion Operator $\Psi$ and its Backward Gradients}
\label{sec:Psi}
We remark that we continue using the physical forward operator for inversion (instead of a neural net parameterised forward operator), and therefore, the analytic inversion can be obtained by solving the first sub-problem of \eqref{lddmm_admm}, which reads:
\begin{equation}
\label{Psi_forward}
\Psi(v, y, \rho)=\left(A^{\top} A+\rho \mathcal{I}\right)^{-1}  (A^{\top} y+\rho v).
\end{equation}

\noindent
One can  numerically solve \eqref{Psi_forward} by conjugate gradient. With this purpose, the derivatives for $\Psi$ can be obtained by differentiating the following expression:
\begin{equation}
\label{eqn:deq}
\left(A^{\top} A+\rho \mathcal{I}\right)\Psi = A^{\top} y+\rho  v,
\end{equation}

\noindent
we then get:
\begin{equation}
\label{eqn:diff}
\left(A^{\top} A+\rho \mathcal{I}\right)\partial \Psi + \Psi \partial \rho = A^{\top} \partial y+\rho   \partial v +  v \partial \rho
\end{equation}

\noindent
Then the derivatives of $\Psi$  are given by:
\begin{equation}
\label{Psi_derivative}
{\partial \Psi}= \rho\left(A^{\top} A+\rho \mathcal{I} \right)^{-1} {\partial v} + \left(A^{\top} A+\rho \mathcal{I} \right)^{-1} A^{\top} {\partial y} +\left(A^{\top} A+\rho \mathcal{I} \right)^{-1}(v - \Psi)\partial \rho.
\end{equation}

To give the backward gradients for the backpropagation algorithm, let  $  f : \mathbb{R}^{ n} \rightarrow \mathbb{R}$ - then  the derivatives of $f(\Psi(v,y,\rho))$ with respect to $v$, $y$ and $\rho$ can be correspondingly computed by:
\begin{equation}
\label{Psi_deri}
\left \{
\begin{array}{ll}
\frac{\partial f(\Psi(v,y,\rho))}{\partial v}=[\rho\left(A^{\top} A+\rho \mathcal{I} \right)^{-1} ]^\top\frac{\partial f}{\partial \Psi }=\rho\left(A^{\top} A+\rho \mathcal{I} \right)^{-1} \frac{\partial f}{\partial \Psi }\\
\frac{\partial f(\Psi(v,y,\rho))}{\partial y}=[\left(A^{\top} A+\rho \mathcal{I} \right)^{-1} A^{\top}]^\top\frac{\partial f}{\partial \Psi }=A \left(A^{\top} A+\rho \mathcal{I} \right)^{-1}  \frac{\partial f}{\partial \Psi }\\
\frac{\partial f(\Psi(v,y,\rho))}{\partial \rho}=[ \left(A^{\top} A+\rho \mathcal{I} \right)^{-1}(v - \Psi)]^\top \frac{\partial f}{\partial \Psi }=(v - \Psi)^\top \left(A^{\top} A+\rho \mathcal{I} \right)^{-1}\frac{\partial f}{\partial \Psi }
\end{array}
\right .
\end{equation}

\noindent
For the inversion  $\left(A^{\top} A+\rho \mathcal{I} \right)^{-1} \frac{\partial f}{\partial \Psi} $, one can compute the derivatives  of $f(\Psi)$ with respect to $v$, $y$ and $\rho$ by applying conjugate gradient.

\subsection{A Deep Registration Net $\Phi$ for Image Shape Prior}\label{sec:Phi}
In this subsection, we establish a neural-network-parameterised registration mapping, which serves as image shape prior for inversion block. Our motivation comes from recent developments on vector momentum-parameterised deep networks proposed, for example, in~ \cite{yang2017quicksilver, shen2019networks}, in which authors showed promising accuracy and significant speedup in obtaining the initial momentum prediction.
With this motivation in mind, in this work, we split the deep registration net $\Phi(t,g,\theta)$ into two-Nets: a momentum prediction net $m=\Lambda(t,g,\theta_1)$ and shooting-warping net $\Gamma(m,g,\theta)$. These nets are applied to each stage $k$. The momentum net is expressed as:

\begin{equation}
m^{k+1}=\Lambda(t^k,g,\theta_{1,k})
\end{equation}

\noindent
whilst the warp Net reads:
\begin{equation}
f^{k+1}=\Gamma(m^{k+1},g,\theta_{2,k}).
\end{equation}

\noindent
That is- it can be expressed as:
 \begin{equation}
 f^{k+1} = \Phi(t^k, g,\theta_k)= \Gamma(\Lambda(t^k,g,\theta_{1,k}),g,\theta_{2,k})
\end{equation}

\begin{figure}[t!]
\centering
\includegraphics[width=1\textwidth]{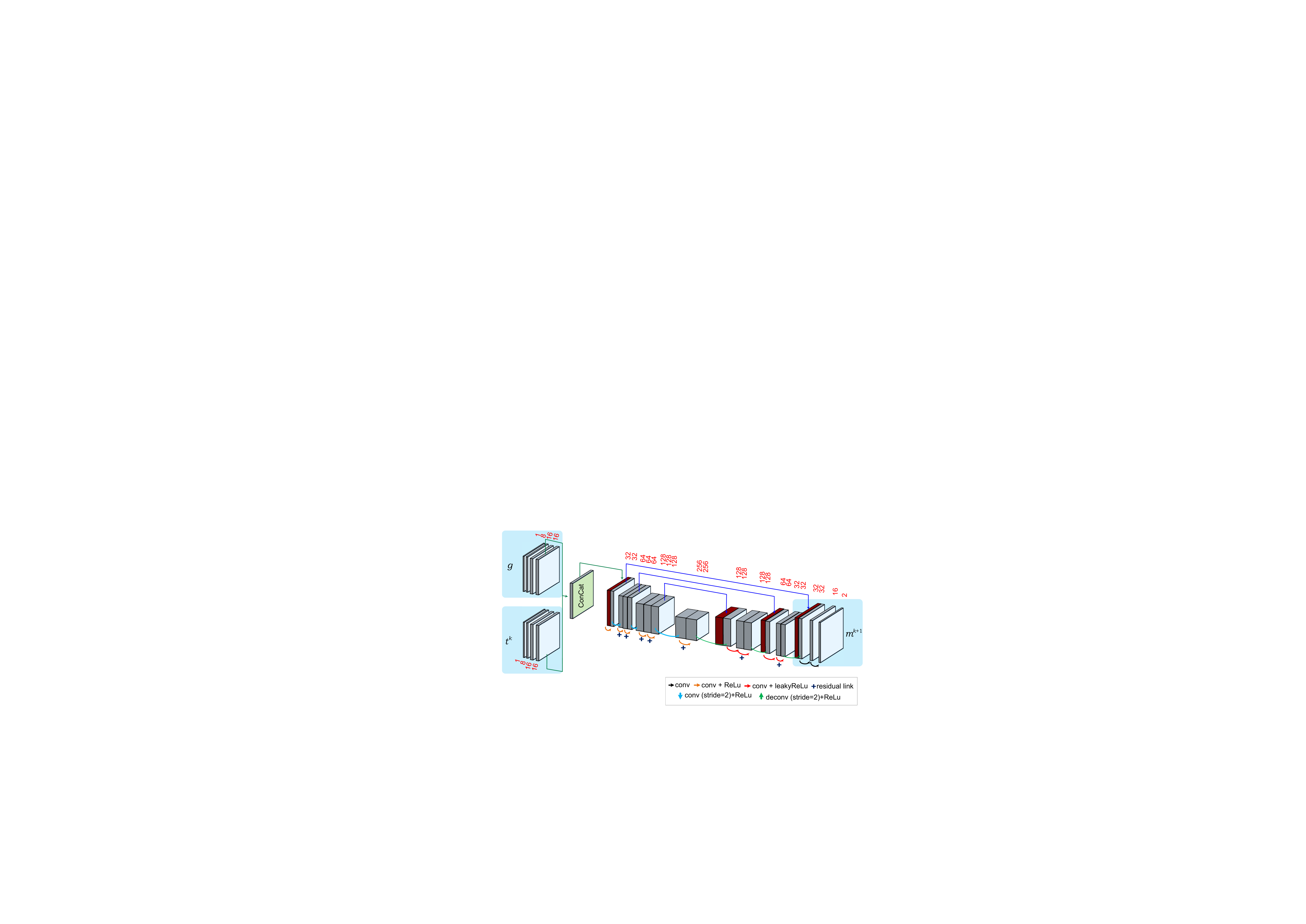}
\caption{\label{fig:momentum} Detailed architecture for the momentum prediction net $\Lambda(t^k,g,\theta_{1,k}) \rightarrow m^{k+1}$ .} \end{figure}

\begin{figure}[t!]
\centering
\includegraphics[width=1\textwidth]{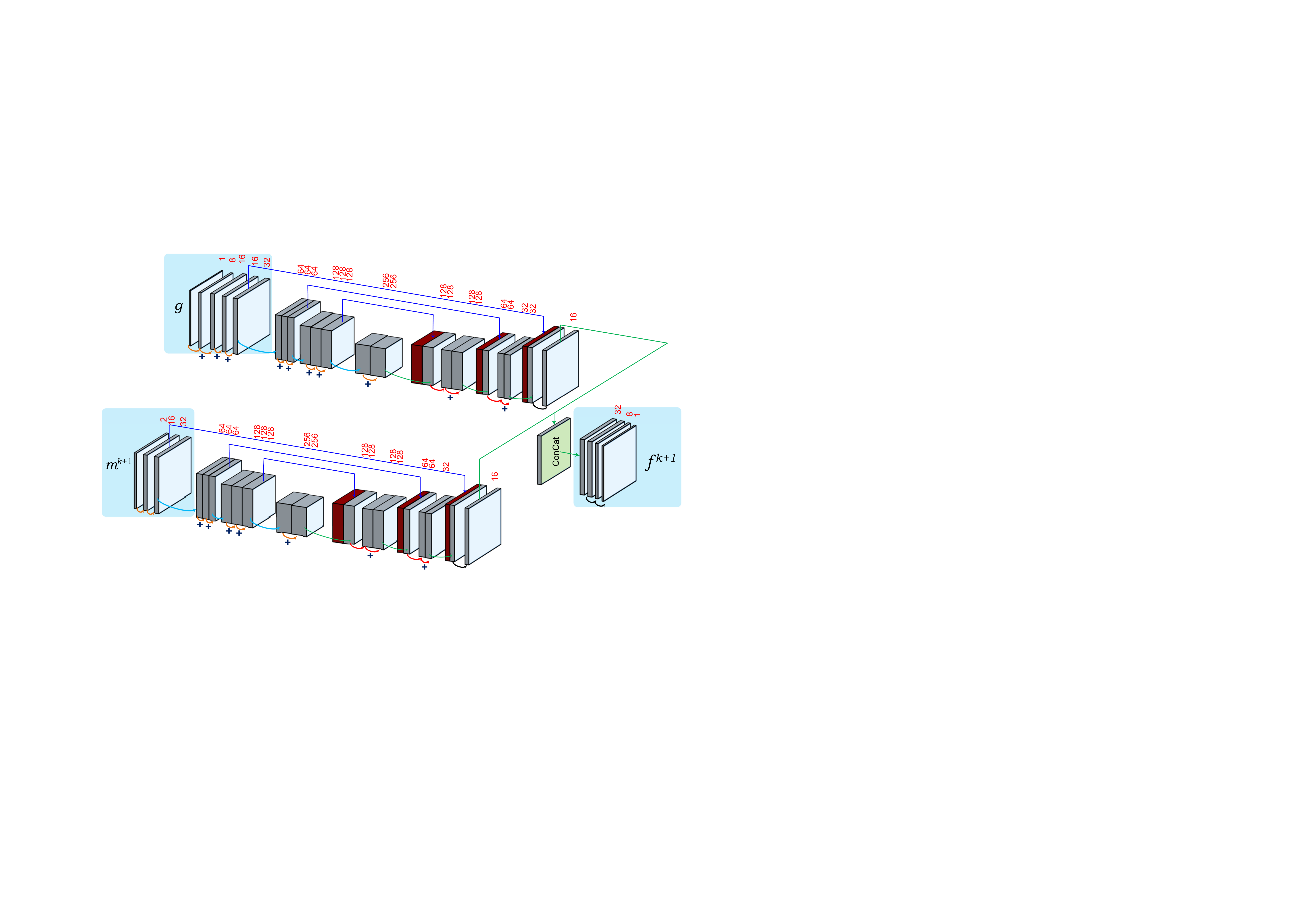} 
\caption{\label{fig:phi}Detailed architecture used for the shooting-warping net $\Gamma(m^k,g,\theta_{2,k})\rightarrow f^{k+1}$.}
\end{figure}

In this work, for the momentum prediction, we use the vector momentum-parameterised stationary velocity field (vSVF) model of that~\cite{shen2019networks}. This is displayed in Fig.~\ref{fig:momentum}. For the Shooting-warping Net $\Gamma$, we propose an extension of the momentum Net to a symmetrical-like Net, whose detailed structure can be seen in Fig. \ref{fig:phi}.

\subsection{Loss function with momentum regularised via LDDMM}
We denote the input template images and acquired measurements as $\{\bm{g}_\ell, \bm{y}_\ell\}_{\ell=1}^{l}$   with corresponding ground truth target images denoted as $\{\bm{f}_\ell \}_{\ell=1}^{l}$. Moreover, let $\Theta$ be the collection of the weights of all registration Nets $\{\theta_k\}_{k=1}^N$.  We then use the following loss function:

\begin{equation}
\label{eqn:lossfun}
 \mathcal{L}(\Theta) =  \mathcal{E}(\Theta) + \mathcal{R}(\Theta),
\end{equation}
where

\begin{equation}\label{eqn:energ}
 \mathcal{E}(\Theta) =  \alpha_N\|\bm{u}^N (\Theta)-\bm{f}\|_2^2,
\end{equation}

\noindent
and $\mathcal{R}(\Theta)$, which seeks to regularise the registration parameters and  guarantees physical meaning of all blocks, is denoted as:

\begin{equation}\label{eqn:thetareg}
 \mathcal{R}(\Theta) = \sum_{i=1}^{N}\alpha_i \|\bm{u}^i -\bm{f}\|_2^2 + \sum_{i=1}^{N}\beta_i \|\bm{m}^i -\tilde{\bm{m}}\|_2^2.
\end{equation}

We remark that, in this work, $\{\tilde{\bm{m}}_\ell\}_{\ell=1}^l$
are obtained from target-template pairs $\{(\bm{f}_\ell, \bm{g}_\ell)\}_{\ell=1}^{l}$ by (\ref{lddmm_shooting}). Therefore, all momentum and reconstructed (warped) images can be obtained simultaneously, in which we seek that they approximate the ground-truth gradually stage by stage. That is,
\begin{equation}\label{eqn:regu}
\|\bm{u}^1   -\bm{f}\|_2^2 \geq \|\bm{u}^2  -\bm{f}\|_2^2 \geq \cdots \geq \|\bm{u}^N  -\bm{f} \|_2^2
\end{equation}
and
\begin{equation}\label{eqn:regm}
\|\bm{m}^1   -\tilde{\bm{m}}\|_2^2 \geq \|\bm{m}^2  -\tilde{\bm{m}}\|_2^2 \geq \cdots \geq \|\bm{m}^N  -\tilde{\bm{m}} \|_2^2.
\end{equation}


After we obtain the learned network parameter set ${\Theta}^*$,  the learned network
\begin{equation}
\label{fixpit_inf}
\begin{split}
t^{k+1}=&t^k-\Gamma(\Lambda(t^k,g,\theta^\ast_{1,k}),g,\theta^\ast_{2,k})+\Psi(2\Gamma(\Lambda(t^k,g,\theta^\ast_{1,k}),g,\theta^\ast_{2,k})-{t}^k, y, \rho_k)), \\
\end{split}
\end{equation}

\noindent
for $k=0, 1, \cdots, N-1$,  is ready to be used for mapping a given measurement-template data pair $(y,g)$ to a predicted  momentum  $m^\ast$ by the output of the last momentum net, that is:
\begin{equation}
\label{predicted-m}
m^\ast=\Lambda(t^N,g,\theta_{2,N}^\ast),
\end{equation}
For estimating   $u^\ast$, one can have two options. As first option, $u^\ast$ can be obtained from the output of the last shooting-warping net  as:

\begin{equation}
\label{estimated-im}
u^\ast=\Phi(t^N,g, \theta_N^\ast).
\end{equation}

Alternatively, the predicted momentum $m_t(x,0)=m^\ast$ can be used to obtain  $\phi(x,-1)$ via the shooting equations:
\begin{equation}
\label{phi_shooting}
\left \{
\begin{array}{l}
  \phi_t(x,t)=v(\phi(x,t),t)\\
   \phi(x,0)=\mathcal{I} \\
   m(x,t)-L v(x,t)=0 \\
   m_t(x,t)+ad^*_{v} m(x,t)=0\\
   m_t(x,0)=m^\ast \\
  \end{array}
\right .
\end{equation}
and finally, as a second option, we can get the estimated ground truth image by:
\begin{equation}
\label{warped-im}
u^\ast = g\circ\phi(x,1).
\end{equation}

\noindent
In the experimental results, we include an ablation study to show the benefits of computing $u^\ast$ using \eqref{estimated-im} and \eqref{warped-im}.

\section{Experimental Results}
In this section, we describe in details the experiments conducted to validate our proposed framework.

\subsection{Data Description}
We remark that whilst our approach can be applied to different medical modalities. In this work, we showcase our approach for MRI, sparse-view CT and low dose CT.

\begin{itemize}
\itemsep0em
    \item \textbf{Dataset A [MRI Dataset]}: Cardiac cine MRI data coming from realistic simulations generated using the MRXCAT phantom framework~\cite{wissmann2014mrxcat}. The heart beat and respiration parameters were set to 1s and 5s respectively. Moreover, the Matrix size is $409 \times 409$, heart phases= 24 and coils=12.
    \item \textbf{Dataset B [Sparse-view CT Dataset]}: We use the Thoracic 4D Computed Tomography (4DCT) dataset~\cite{castillo2009four}\footnote{https://www.dir-lab.com/Downloads.html}. The measurements are generated by: $y=Au$  with 18 views over $360^{\circ}$, where $A$ is X-ray transform and $u$ is normalised to $[0,1]$.
    \item \textbf{Dataset C [Low Dose CT Dataset]}: As in Dataset B we use Thoracic 4D Computed Tomography (4DCT) dataset~\cite{castillo2009four}. However, the measurements are generated by: $y=A(u+\sigma\xi)$ with 181 views over $360^{\circ}$ and $\xi$ obey i.i.d  normal distribution, $\sigma =0.10$.
\end{itemize}

We remark that the MRI measurements are generated  by partial Fourier transform as: $y=K \mathcal{F}(u+\sigma(\xi_1+\xi_2*i))$. Where $\sigma$ is the  noise level, $\xi_1, \xi_2$ obey i.i.d  normal distribution,  $u$ is the ground truth image, and $K$ is  the undersampled operator, and $\mathcal{F}$ is Fourier Transform. In this work, we retrospectively undersampled the measurements using: radial sampling, 2D random variable-density  with fully sampled center radius and 1D variable-density with fully sampled center. To show generalisation capabilities of our proposed approach, we ran our approach using different sampling rates = $\{1/5, 1/4, 1/3 \}$

\subsection{Parameter Selection and Setting Details}
In this part, we give further details on the choice of the parameters along with further specifics of how we ran our experimental results.

\medskip
For the $\Psi$ and  $\Phi$ Nets, we set the  number of stages $N = 3$ for
all our applications: for fast MRI , sparse-view CT, and low-dose CT. Our approach is a GPU-based  implementation  in  Pytorch.  The $\rho$ in  $\Psi$ are set to be learnable,  and we also restrict $\rho\in [0,c]$ by adding a layer as: $\rho= c\sigma(0.4w)$,
where $\sigma = \frac{e^x}{1-e^x}$ is a Sigmoid function, $w$ is learnable, and $c=0.8$  to prevent  $\rho$ to become too big.

We use Adam algorithm for training with the following parameters: learning rate: 1e-4, epochs= 500. Moreover, for the  learned $\rho:$ MRI
$\rho= [0.16,0.26, 0.33]$; sparse-view CT $\rho= [0.55,0.34, 0.41]$ and low-dose CT $\rho= [0.64,0.42, 0.38]$

\medskip
\textbf{Setting for the MRI Case.} The temporal cine cardiac data (Dataset A) is  used to generate  376 2D image pairs as target-template image pairs, and then the momentums dataset associated with target-template image pairs is obtained via LDDMM (\ref{lddmm_shooting}) for regularising the momentum prediction Nets in our approach \eqref{eqn:fixpit-net}. In this work,  $u$ is normalised to $[0, 1]$ and set noise level $\sigma=0.05$. We use undersampling rate of $\{1/5, 1/4, 1/3\}$.  In each experiment, 360  measurement-template  pairs with 360  target images and 360 momentums are  used to train our proposed approach (\ref{eqn:fixpit-net}), and 16  measurement-template  pairs are used for testing by (\ref{fixpit_inf}). For speedup the training, we pretrain the model stage by stage for 500 epoch, and finally train the whole network for 500 epochs.

\medskip
\textbf{Setting for the Sparse-view and Low-dose CT Case.}  We generate 528 2D image pairs as target-template,
and then the momentum is obtained via LDDMM (\ref{lddmm_shooting}) for regularising the momentum prediction Net. We use for the Randon Transform $A$ the CUDA version of~\cite{gao2012fast}.
For the training  the network (\ref{eqn:fixpit-net}), 480  measurement-template  pairs with 480  target images, and 480 momentum are  used. Whilst for testing (\ref{fixpit_inf}),  48  measurement-template  pairs are used.

\subsection{Evaluation Methodology} \label{sub:EvalProt}
We evaluate our proposed framework based on the following scheme.

\medskip
\textbf{Comparison against other MRI reconstruction schemes.} For the first part of our evaluation, we compared our framework against the well-established compressed sensing (CS) reconstruction scheme. We solve the CS scheme with TV, and LDDMM computed sequentially. Furthermore, we ran experiments using three different sampling patterns:  radial, 2D random and 1D random (cartesian). To show generalisation capabilities, we use different sampling rates = \{1/5,1/4, 1/3 \}.

We report the results of these comparisons based on both qualitative and quantitative results. The former is based on visual assessment of the reconstruction, and the latter on the computation of two well-established metrics:  the  structural  similarity  (SSIM)  index and the  Peak   Signal-to-Noise   Ratio   (PSNR); along with the computational cost given in seconds.

\textbf{Generalisation capabilities using CT data.} For generalisation capabilities, we evaluate our framework using data coming from sparse view CT and low-dose CT. We compared our framework against classic TV-reconstruction scheme + LDDMM computed sequentially and another indirect registration approach that of ~\cite{chen2018indirect}. We report the comparison using qualitative and quantitative results using  visual comparison of the reconstructions along with the error maps, reconstruction quality in terms of PSNR, SSIM and computation cost.

\subsection{Results and Discussion}
In this subsection, we demonstrate the capabilities of our framework following the evaluation scheme of subsection~\ref{sub:EvalProt}.

\medskip
\textbf{$\triangleright$ Is Our Framework better than a classic MRI Reconstruction Scheme?} We begin by evaluating our approach against classic TV+LDDMM reconstruction scheme. We remark to the reader that classic scheme performs sequentially the reconstruction and registration whilst our approach computes simultaneously the MRI reconstruction and indirect image registration.

\begin{figure}[H]
\centering
\includegraphics[width=0.88\textwidth]{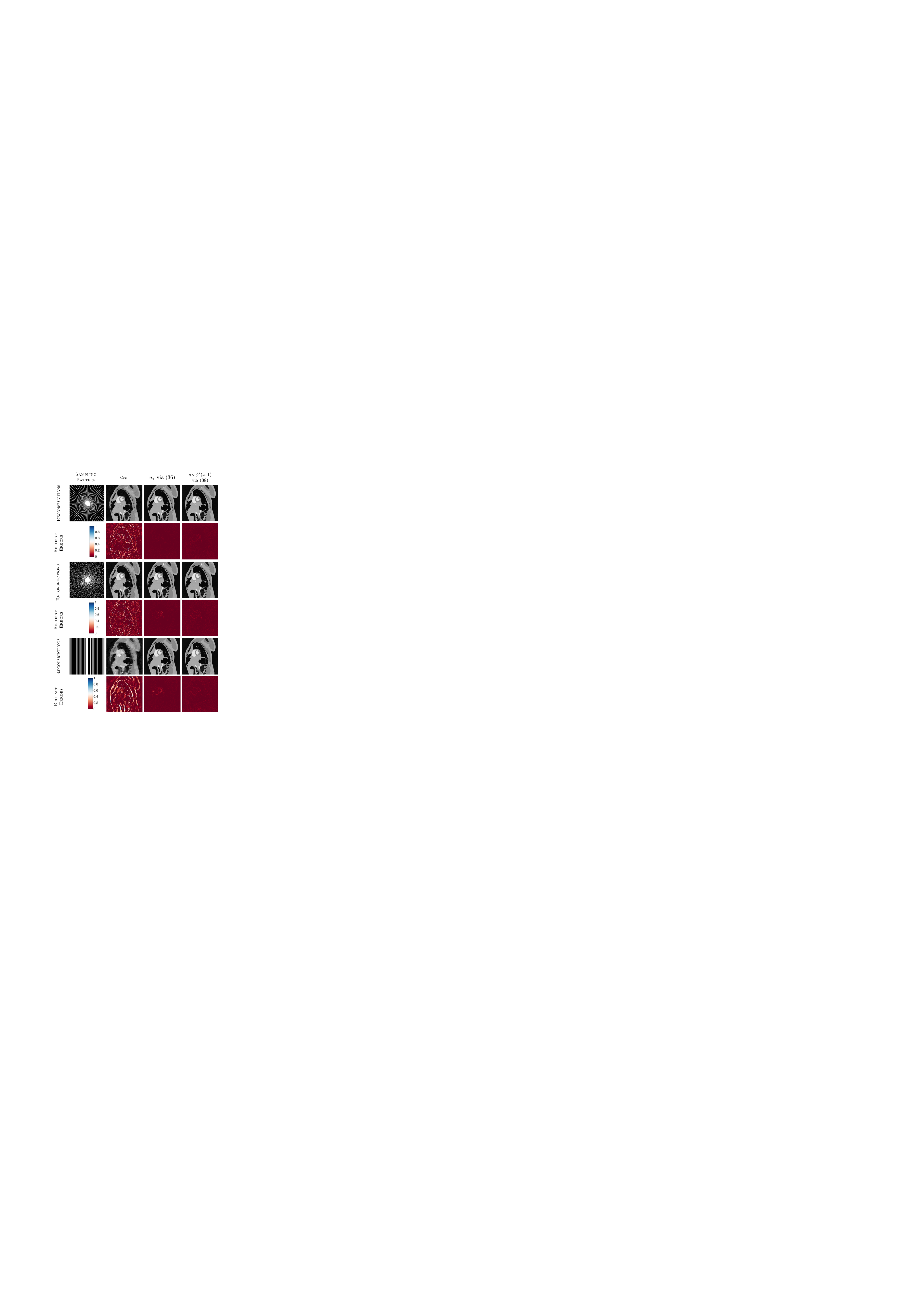} \vspace{-0.55cm}
\caption{MRI Reconstruction outputs and reconstruction errors using Dataset A with sampling rate = $1/5$.
Comparison of our approach vs classic scheme  (TV + LDDMM). Our approach reconstruct higher quality images with sharp edges, preservation of fine details and contrast.}
\label{fig::20Sam}
\end{figure}

\begin{figure}[H]
\centering
\includegraphics[width=0.88\textwidth]{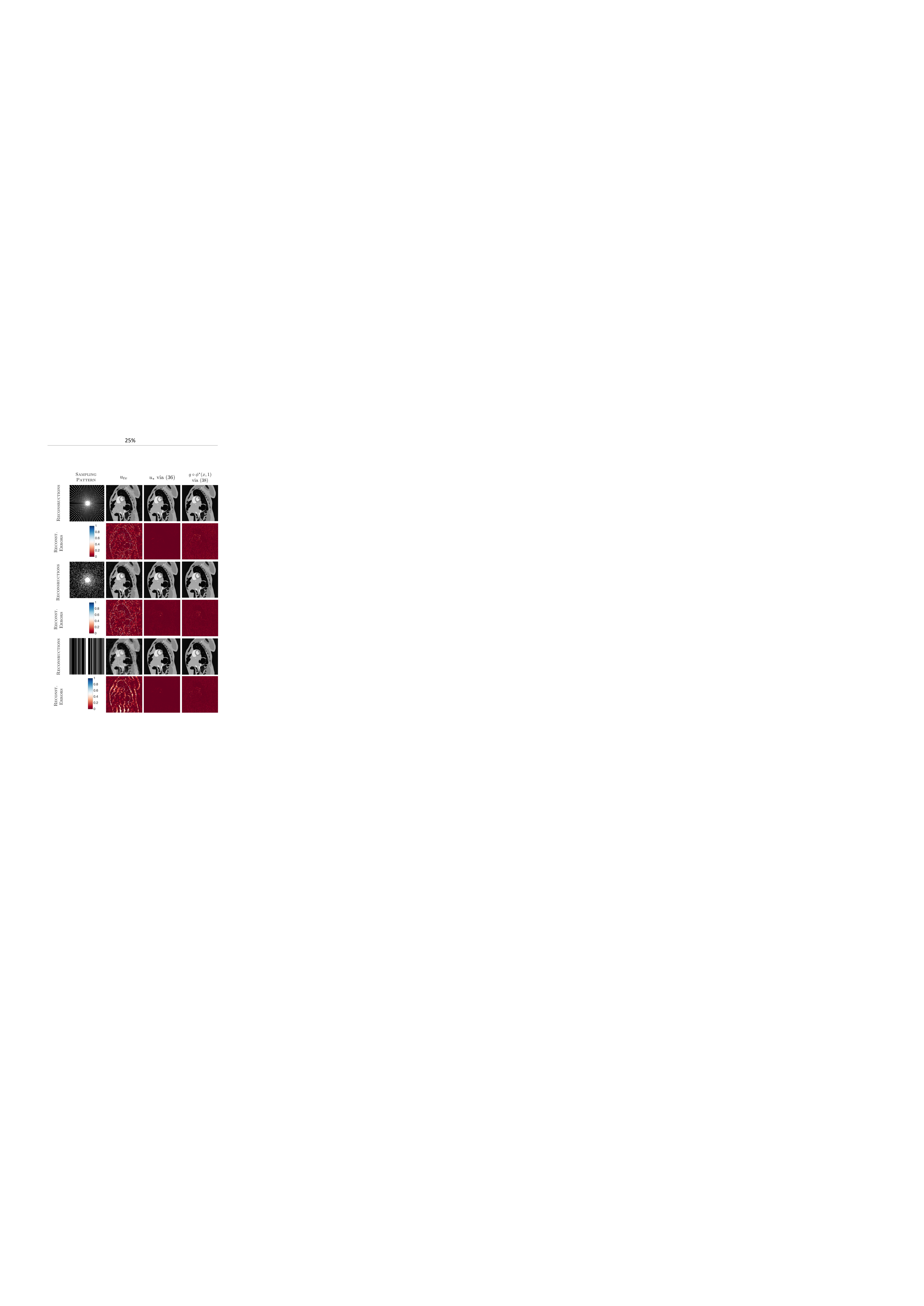} \vspace{-0.55cm}
\caption{MRI Reconstruction outputs and reconstruction errors using Dataset A with sampling rate = $1/4$ and with different sampling patterns. Results from classic scheme (TV+LDDMM) vs our approach. One can see that our reconstructions have higher quality, this is reflected in the reconstruction error plots.}
\label{fig::25Sam}
\end{figure}

\begin{figure}[H]
\centering
\includegraphics[width=0.88\textwidth]{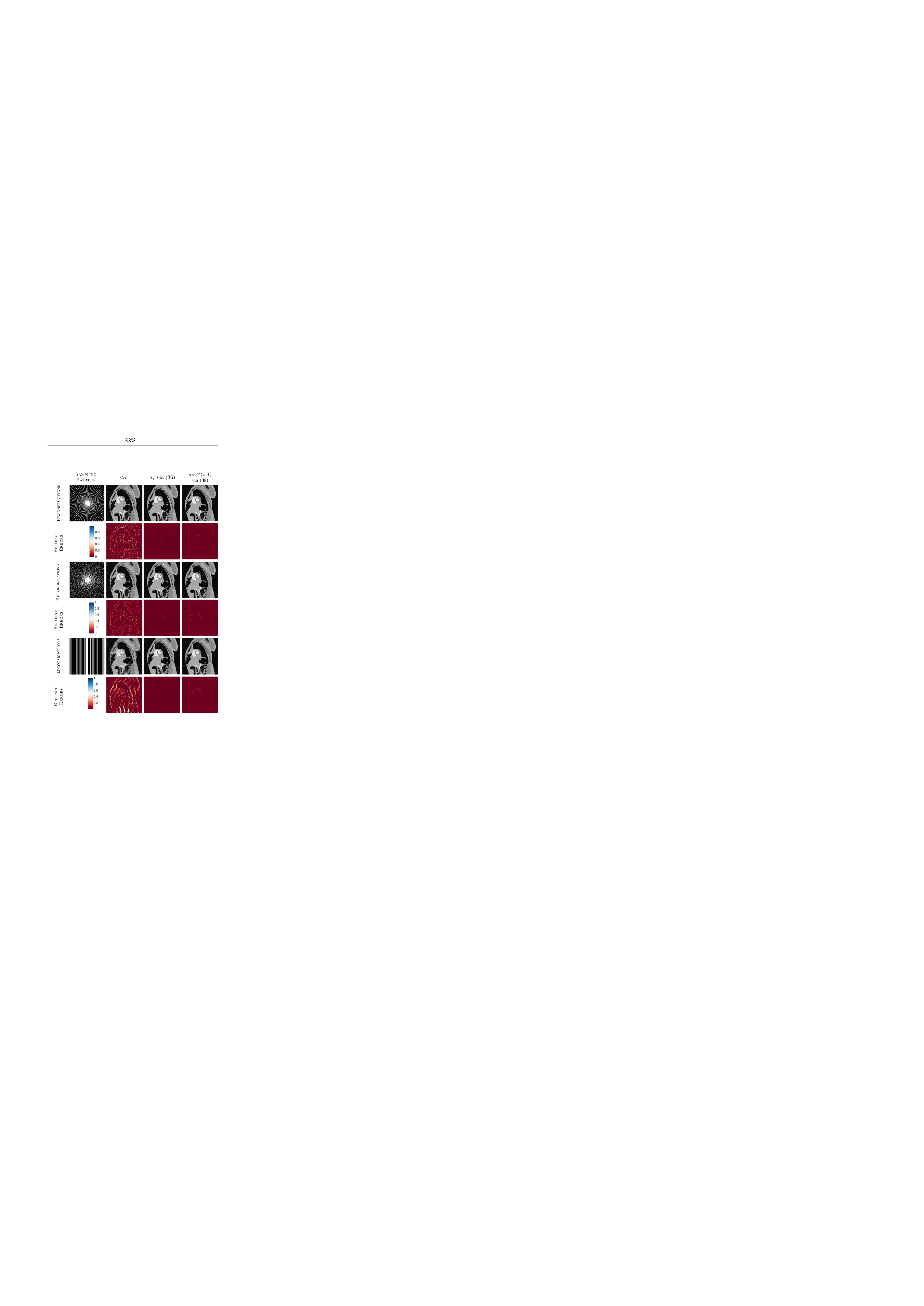} \vspace{-0.55cm}
\caption{MRI Reconstruction outputs and reconstruction errors using Dataset A with sampling rate = $1/3$ and with different sampling patterns. Reconstructions show that our approach reconstructs higher quality images than classic scheme TV+LDDMM. This is further supported by the reconstruction error plots, in which our reconstructions reported the lowest error.}
\label{fig::30Sam}
\end{figure}

We report both qualitative and quantitative results in Table \ref{tab::MRIresults} and Figs.  \ref{fig::20Sam}, \ref{fig::25Sam} and \ref{fig::30Sam}. In Fig.~\ref{fig::20Sam}, we show nine reconstructed output examples  with three different sampling patters. Visual assessment agrees with the theory of our model, in which we highlight the reconstruction of higher quality and preservation of relevant anatomical parts whilst enhancing  fine details and contrast. In a closer inspection at these reconstructions, one can see that our framework (in both cases either using ~\eqref{estimated-im} or  ~\eqref{warped-im}) leads to  reconstructions with sharper edges and better preservation of fine details than the classic MRI reconstruction scheme.  This is further supported by the reported reconstruction errors, in which our approach reported the lowest error values for all reconstructed samples.

\begin{table}[t!]
\small
\centering
\caption{Numerical comparison of our approach vs. other reconstruction schemes using the Dataset A, with different reconstruction patterns and acceleration factors. Results are reported from the testing set.  SSIM is denoted in $10^{-2}$. \crule[green]{0.3cm}{0.3cm} denotes the best image quality scores whilst   \crule[yellow]{0.3cm}{0.3cm} the lowest computational cost.}\label{tab:cmp_mri}
\begin{tabular}{c|c|c|c|c}
\hline
\cellcolor[HTML]{EFEFEF} \textsc{Pattern} & \cellcolor[HTML]{EFEFEF}\textsc{Quantity}  & \cellcolor[HTML]{EFEFEF}\textsc{TV+LDDMM}  & \cellcolor[HTML]{EFEFEF}\textsc{\textcolor{blue}{Ours}}~\eqref{estimated-im} &  \cellcolor[HTML]{EFEFEF}\textsc{\textcolor{blue}{Ours}}~\eqref{warped-im} \\
\hline \hline
\multicolumn{5}{c}{\textsc{Dataset A  with Sampling Rate = 1/5 }} \\ \hline
\multirow{2}{*}{\textsc{Radial}}
& (PSNR, SSIM) &     (25.84, 77.36) & \cellcolor[HTML]{9AFF99}(37.90, 93.59) & \cellcolor[HTML]{9AFF99}(35.11, 88.25) \\
\cline{2-5}
& Time Cost (s) &   1.54   & \cellcolor[HTML]{FFFFC7} 0.52 & 0.61  \\
\hline
\multirow{2}{*}{\textsc{2D Random}}
& (PSNR, SSIM) &  (25.06, 77.61) & \cellcolor[HTML]{9AFF99}(36.08, 93.34) & \cellcolor[HTML]{9AFF99}(34.32, 88.38) \\
\cline{2-5}
& Time Cost (s) &     1.66   &  \cellcolor[HTML]{FFFFC7} 0.56 & 0.67  \\
\hline
\multirow{2}{*}{\textsc{1D Random}}
& (PSNR, SSIM) &  (20.61, 61.31) & \cellcolor[HTML]{9AFF99}(36.10, 93.31) & \cellcolor[HTML]{9AFF99}(34.99, 88.42)\\
\cline{2-5}
&  Time Cost (s)&     1.51   &  \cellcolor[HTML]{FFFFC7} 0.51 & 0.63  \\
\hline \hline
\multicolumn{5}{c}{\textsc{Dataset A  with Sampling Rate = 1/4 }} \\ \hline
\multirow{2}{*}{\textsc{Radial}}
& (PSNR, SSIM) &     (26.52, 78.89) & \cellcolor[HTML]{9AFF99}(38.77, 94.43) & \cellcolor[HTML]{9AFF99}(35.74, 90.18) \\
\cline{2-5}
& Time Cost (s) &  1.60   & \cellcolor[HTML]{FFFFC7} 0.57  &  0.63 \\
\hline
\multirow{2}{*}{\textsc{2D Random}}
& (PSNR, SSIM) &  (25.94, 78.19) & \cellcolor[HTML]{9AFF99}(38.12, 94.42) & \cellcolor[HTML]{9AFF99}(35.70, 90.44) \\   
\cline{2-5}
& Time Cost (s) & 1.63      & \cellcolor[HTML]{FFFFC7}  0.53 & 0.71  \\
\hline
\multirow{2}{*}{\textsc{1D Random}}
& (PSNR, SSIM) &  (22.02, 65.67) & \cellcolor[HTML]{9AFF99}(37.44, 94.33) & \cellcolor[HTML]{9AFF99}(35.82, 90.18)\\
\cline{2-5}
&  Time Cost (s)&   1.58     & \cellcolor[HTML]{FFFFC7} 0.56  &  0.66 \\
\hline \hline
\multicolumn{5}{c}{\textsc{Dataset A  with Sampling Rate = 1/3 }} \\ \hline
\multirow{2}{*}{\textsc{Radial}}
& (PSNR, SSIM) &     (26.82, 79.63) & \cellcolor[HTML]{9AFF99}(39.01, 94.63) & \cellcolor[HTML]{9AFF99}(35.77, 90.36) \\
\cline{2-5}
& Time Cost (s) &    1.57   & \cellcolor[HTML]{FFFFC7} 0.56  &  0.64 \\
\hline
\multirow{2}{*}{\textsc{2D Random}}
& (PSNR, SSIM) &  (26.18, 78.77) & \cellcolor[HTML]{9AFF99}(38.79, 94.75) & \cellcolor[HTML]{9AFF99}(35.78, 90.65)\\
\cline{2-5}
& Time Cost (s) &      1.47   & \cellcolor[HTML]{FFFFC7} 0.49  &  0.63   \\
\hline
\multirow{2}{*}{\textsc{1D Random}}
& (PSNR, SSIM) &  (22.60, 66.83) & \cellcolor[HTML]{9AFF99}(38.45, 94.42) & \cellcolor[HTML]{9AFF99}(35.84, 90.21)\\
\cline{2-5}
&  Time Cost (s)&     1.64   & \cellcolor[HTML]{FFFFC7} 0.56  &  0.59  \\ \hline
\end{tabular}
\label{tab::MRIresults}
\end{table}

To show further generalisation capabilities, we ran a range of experiments using different sampling factors = \{1/5, 1/4, 1/3\}. Reconstruction outputs can be seen in Figs. \ref{fig::20Sam}, \ref{fig::25Sam} and \ref{fig::30Sam}. One can see that the benefits of our approach described above are prevalent to all  sampling factors. That is, our approach preserves small structures for example the  papillary  muscles  of  the  heart. Moreover, in a visual comparison between these figures, we notice that our method generalises very well even when the acceleration factor is increasing; contrary to the classic scheme that exhibits loss of contrast and blurry effects. Overall, we can show that providing a shape prior, through a registration task, yields to higher quality images whilst decreasing
the number of measurements to form an MRI.

\begin{table}[t!]
	\centering
	\caption{\small{Numerical comparison for sparse-view and low-dose CT datasets (B\&C). The displayed results are the averaged accuracy and efficiency on the testing dataset. \crule[green]{0.3cm}{0.3cm} denotes the best image quality scores whilst   \crule[yellow]{0.3cm}{0.3cm} the lowest computational cost. }}
	{\small
	 \resizebox{\textwidth}{!}{
		\begin{tabular}{c|c|c|c|c}
			\hline
			\cellcolor[HTML]{EFEFEF}\textsc{Quantity} 	& \cellcolor[HTML]{EFEFEF}\textsc{TV+LDDMM} & \cellcolor[HTML]{EFEFEF}\textsc{Chen et al.}~\cite{chen2018indirect} & \cellcolor[HTML]{EFEFEF}\textsc{\textcolor{blue}{Ours}} (\ref{estimated-im}) & \cellcolor[HTML]{EFEFEF}\textsc{\textcolor{blue}{Ours}} (\ref{warped-im})\\
			\hline \hline
			\multicolumn{5}{c}{\textsc{Dataset B}} \\ \hline
			 (PSNR,SSIM) &  (26.71, 0.72) & (30.11, 0.96) & \cellcolor[HTML]{9AFF99}(36.34, 0.97) & \cellcolor[HTML]{9AFF99}(34.48, 0.95)\\
			\hline
			 Time Cost (s) &  1.82 &  81.37 & \cellcolor[HTML]{FFFFC7} 0.76 & 0.87 \\
			\hline \hline
			\multicolumn{5}{c}{\textsc{Dataset C}} \\ \hline
            (PSNR, SSIM) & (30.66, 0.86)&  (31.41, 0.95) & \cellcolor[HTML]{9AFF99}(39.18, 0.97) & \cellcolor[HTML]{9AFF99}(35.78, 0.96)\\
            \hline
            Time Cost (s) &  1.73 & 112.35 & \cellcolor[HTML]{FFFFC7} 0.84 & 1.08\\
			\hline
		\end{tabular}}
	}
\label{tab:CTrec}
\end{table}

\smallskip
\textbf{$\triangleright$ Is a Two-task Model better than a Sequeantial Model - Does It Pay Off?} To further support the aforementioned benefits of our model and for a more detailed quantitative analyses, we report the overall results of the Dataset A in Table~\ref{tab::MRIresults}. The results are the average of the image metrics, (PSNR, SSIM), across the whole Dataset A with different sampling patterns and sampling rates.   We observe that our approach reported significant improvement in both metrics with respect to the classic MRI + LDDMM reconstructions and for all accelerations. These results further validate our hypothesis that providing shape prior improve substantially the reconstruction image quality.

After demonstrating the benefits of our approach quality-wise, we now pose a question- how is our approach performing from a computational point of view? The computational time is displayed in  Table~\ref{tab::MRIresults}. One can observe that another major advantage of our model is the computational time, we achieve to decrease an average of ~65\% the computation cost with respect to the classic reconstruction scheme whilst achieving a substantial improvement in terms of image quality in both metrics. Overall, the potentials of our approach are preserved for all datasets and for all sampling rates.

\begin{figure}[t!]
\centering
\includegraphics[width=1\textwidth]{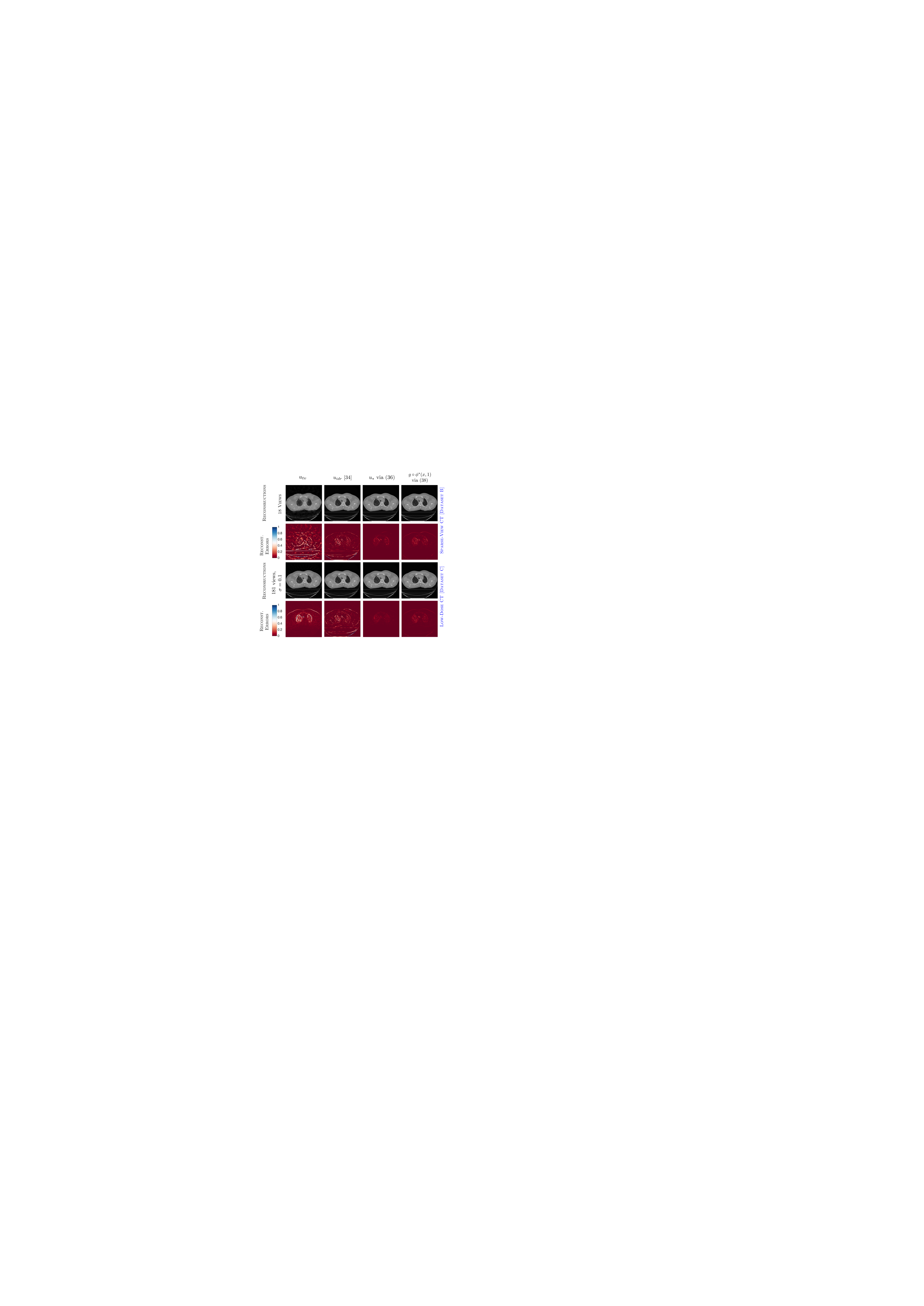} \vspace{-0.8cm}
\caption{CT reconstruction outputs and reconstruction errors using Datasets B and C. A comparison is displayed between classic reconstruction scheme and our approach. In a closer inspection, one can see that our reconstructions have higher image quality than the compared schemes. This is further supported by the reconstruction error plots, in which our reconstructions display the lowest errors.}
\label{fig::CTrec}
\end{figure}

\smallskip
\textbf{$\triangleright$ Can Our Approach be Applied to other Modalities? Generalisation Capabilities}
To demonstrate generalisation capabilities of our model, we run experiments on both sparse-view and low-dose  CT datasets (e.g. Datasets B and C). We remark to the reader, that to the best of our knowledge, this is the first hybrid approach reported that performs two tasks as a \textit{hybrid model}. That is- an approach that combines a model-based and a deep learning-based models to improve image reconstruction. However, there is a  model-based approach that follows similar philosophy than ours, which is  that of Chen et al.~\cite{chen2018indirect} that is applied to the CT case. Therefore, we ran our approach and compared against both the classic CT reconstruction scheme with TV + LDDMM, and  that of~\cite{chen2018indirect}.

\begin{figure}[t!]
\centering
\includegraphics[width=1\textwidth]{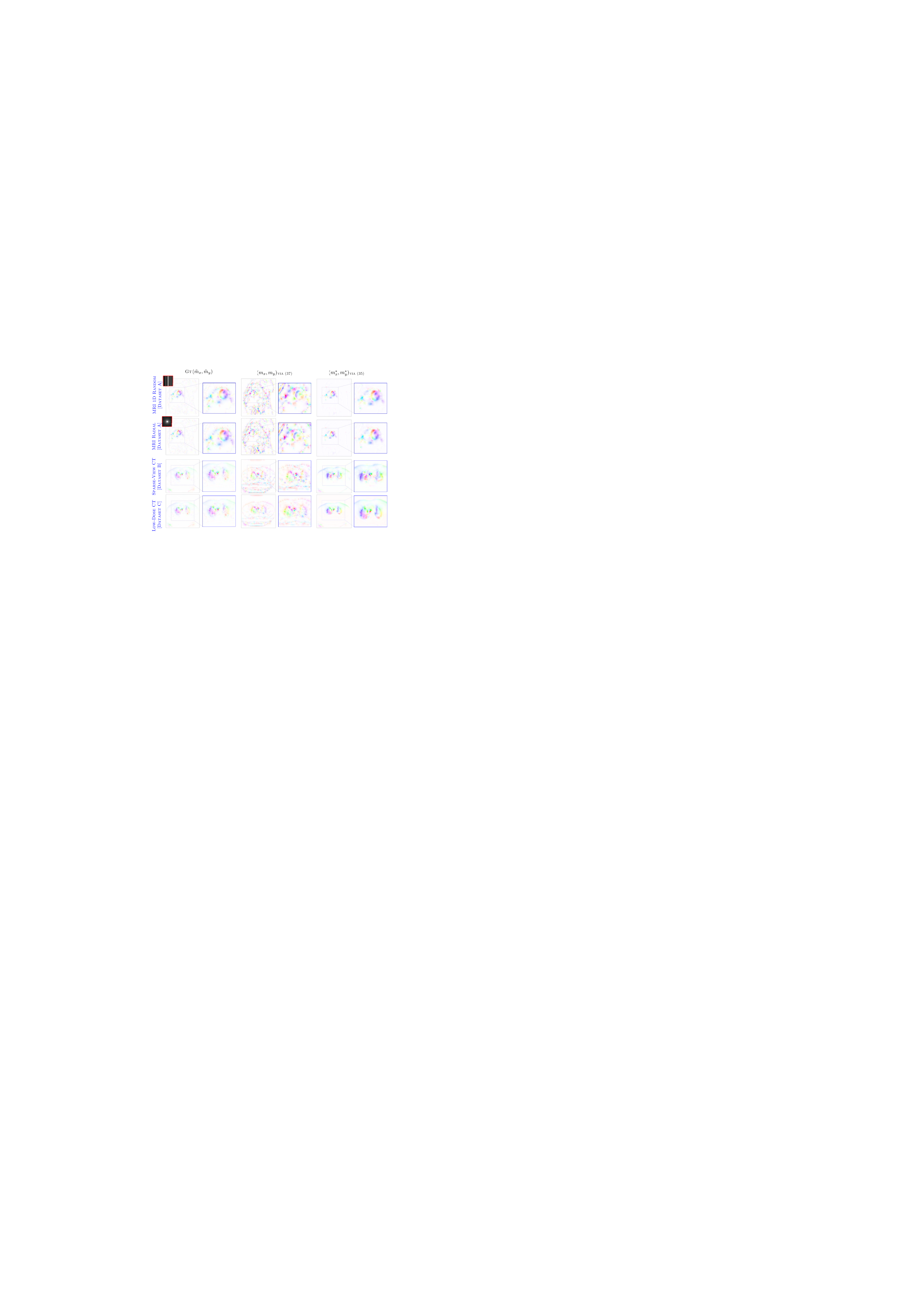} 
\caption{Visualisation of the predicted momentum. (From left to right) ground truth $(\tilde{m}_x, \tilde{m}_y)$ and predicted ones using cartesian and radial sampling patters, and DAtasets A, B and C. }
\label{fig::predictedMomementu}
\end{figure}

We begin by evaluating visually our approach against the compared schemes and the results are displayed in Fig. \ref{fig::CTrec}.  In that figure, we display two samples outputs using datasets B and C respectively.  In a closer look at the reconstructions, one can see that classic TV + LDDMM reconstructions fail to preserve fine details and introduce strong blurring artefacts (see first column). Similarly, the algorithmic approach of that~\cite{chen2018indirect} shows reconstructions with loss in contrast and texture, blurry artefacts and fine  details. These negative effects are reflected at the reconstruction error plots in which our reconstructions (last two columns) reported the lowest errors. From these plots, one can see that our approach is able to reconstruct sharp edges whilst keeping fine details and texture.

To further support our approach, we perform further quantitative experiments, which are reported in Table~\ref{tab:CTrec}.  Similarity-wise we reported the highest values for both  PSNR and SSIM metrics. In particular, we would like to highlight two major potentials of our approach. Firstly, our approach offers substantial improvement, in terms of both image quality metrics. In particular, for the PSNR metric the improvement is highly substantial compared to the approach. Also, in terms of SSIM, it outperforms the classic TV scheme and readily competes against~\cite{chen2018indirect}. Secondly, the computational cost is  significantly lower than the approach of~\cite{chen2018indirect} and the classic reconstruction scheme.  Finally, for further visualisation support, we display the predicted momentum in Fig.~\ref{fig::predictedMomementu}.

\section{Conclusion}
In this paper, we propose for the first time a hybrid approach for simultaneous reconstruction and indirect registration. We demonstrated that indirect image registration, in combination with deep learning, is a promising technique for providing a shape prior to substantially improve  image reconstruction. We show that our framework can significantly decrease the computational cost via deep nets.

In particular, we highlight the potentials  of leveraging physics-driven regularisation methods with the powerful performance  of deep learning in an unified framework. We show that our approach improves over existing regularisation methods. These improvements are in terms of getting higher quality images that preserve relevant anatomical parts whilst avoiding geometric distortions, and loss of fine details and contrast.
Moreover, we also showed that our framework can substantially decrease the computational time by more than ~66\% whilst reporting the highest image quality metrics. These benefits are consistent over different settings such as acceleration factors, sampling patterns and medical image modalities.


\section*{Acknowledgments}
AIAI gratefully acknowledges the financial support of  the  CMIH,  University  of  Cambridge; and Noemie Debroux  for very helpful discussions. \smallskip

CBS acknowledges: inspiring and fruitful discussions with Ozan \"{O}ktem on the topic of indirect image reconstruction and learned image registration, support from the Leverhulme Trust project “Breaking the nonconvexity barrier”, the Philip Leverhulme Prize, the EPSRC  EP/M00483X/1 and EP/S026045/1, the EPSRC Centre EP/N014588/1, the European Union Horizon 2020 research and innovation programmes under the Marie Skodowska-Curie grant agreement No. 777826 NoMADS and No. 691070 CHiPS, the Cantab Capital Institute for the Mathematics of Information and the Alan Turing Institute.


\bibliography{mainSub}


\end{document}